\def\msun{\hbox{$M_\odot$}}
\catcode`\@=11
\def\gsim{\ifmmode{\mathrel{\mathpalette\@versim>}}
    \else{$\mathrel{\mathpalette\@versim>}$}\fi}
\def\lsim{\ifmmode{\mathrel{\mathpalette\@versim<}}
    \else{$\mathrel{\mathpalette\@versim<}$}\fi}
\def\@versim#1#2{\lower 2.9truept \vbox{\baselineskip 0pt \lineskip 
    0.5truept \ialign{$\m@th#1\hfil##\hfil$\crcr#2\crcr\sim\crcr}}}
\catcode`\@=12
\documentclass{iau}
\usepackage{graphicx}

\title{IAUS295 -- The Intriguing Life of Massive Galaxies: Introducing the Final Discussion}

\author[]{Alvio Renzini$^1$}
\affiliation{$^1$INAF -- Osservatorio Astronomico di Padova \\ email: {\tt alvio.renzini@oapd.inaf.it} \\[\affilskip]
}

\pubyear{2013}
\volume{295}  %% insert here IAU Symposium No.
\pagerange{xx--xx}
\jname{The intriguing life of massive galaxies}
\editors{D. Thomas, A. Pasquali \& I. Ferreras, eds.}
\begin{document}

\maketitle

\begin{abstract}
  This is a brief introduction to the closing discussion of the IAU
  Symposium 295, ``The Intriguing Life of Massive Galaxies", that was
  held in Beijing from August 27 through 31, 2012. The discussion was
  focused on only four hot items, namely 1) the redshift evolution of
  the size of passively evolving galaxies, 2) the evolution with
  redshift of the specific star formation rate, 3) quenching of star
  formation in galaxies and dry merging, and 4) the IMF.

\end{abstract}

\firstsection % if your document starts with a section,
\section{Introduction}
I have been asked by the organizers to introduce the final discussion
of this IAU Symposium 295 ``The Intriguing Life of Massive Galaxies"
thus trying to {\it provoke} a wide participation by the audience. I have chosen to focus the discussion 
on just a few among the many issues that
have been addressed and debated at this exciting meeting.  For each topic, after
five minutes of introduction, some ten or more minutes of general
discussion followed. All this was recorded but at this point it is not clear
whether it will become accessible on line on the site of the symposium. Therefore, this short article is limited to a report
of my introduction to each topic, in the wish of attracting readers
towards the actual event, if the records will become available.

I wish to start by confessing that already quite some time ago I lost
faith in the ability of theoretical models to tell us how real
galaxies form and evolve, being them of the semianalytic or the
hydro-simulation variety. The reason is that once the mess of baryonic
physics is added to the clean elegance of dark matter N-body
realizations most --if not all-- predictive power is lost in a
plethora of adjustable parameters meant to describe the many physical
processes at work.  Just to name a few: star formation, galactic
winds, cold streams, supermassive black hole formation, nuclear
activity and its feedback, chemical evolution, galaxy mergers,
starbursts, disk instabilities, multiphase ISM, supernova feedback, ram pressure
stripping, dust formation, radiative transfer and many more.  For example, 
Benson \& Bower (2010) list over two dozens 
adjustable parameter for their semi-analytic model. So,
being an infidel, my view of theory would be biased and I preferred to
drive the discussion towards some specific points that direct observations
may help to clarify.

\section{The Size Evolution of Passively Evolving Galaxies}

Over a dozen talks have been dealing with the size of passively
evolving galaxies (PEG) at the various redshifts. Sizes are
straightforward to visualize and effective radii ($R_{\rm e}$) are relatively easy
to measure, thus over one hundred papers have been dedicated in recent
years to report on how high redshift PEGs are smaller and denser at
given stellar mass compared to their local analogs, and what processes
may account for their apparent growth by up to a factor of $\sim 4$
from $z\sim 2.5$ to $z\sim 0$. Indeed, at fixed stellar mass the
average $R_{\rm e}$ of PEGs  ($<\! R_{\rm e}\!>$, as normalized to the 
local $R_{\rm e}-M_*$ relation) steadily declines with increasing
redshift, a trend that --with few exceptions (e.g., Valentinuzzi et
al. 2010; Cassata et al. 2011; Newman et al. 2012)-- has been often entirely ascribed
to the physical growth of individual galaxies. Accretion of an
envelope of small satellites (minor mergers) has been widely
entertained as the leading mechanism to {\it puff up} the individual
high-$z$ compact PEGs until they reach their due, final
dimension. Yet, this is only part of the story, and likely only a
minor one.

The fact is that the comoving number density of massive PEGs is
increasing by a large factor $(\sim 25$) between $z=2.5$ and 0, with
most of this increase taking place between $z=2.5$ and $\sim 1$. Hence,
the run of the average size of PEGs primarily reflects the size
with which at each redshifts they first appear as quenched, rather
than the subsequent grow of each of them. Thus, the real challenge is to
understand why galaxies which are quenched at later times are born
bigger than those formed earlier, rather than (or not only) to
understand to which extent individual PEGs secularly increase their
size. Precursors to PEGs must be star forming galaxies likely spending
most of their time on their {\it main sequence} growing inside-out.
Hence, the later they are quenched the bigger they are, and presumably the bigger
their passive remnant. Is just this the main story?

An intriguing aspect is that the number density of compact PEGs
actually appears to {\it increase} with cosmic time, peaking at $z\sim
1$ (e.g., Cassata et al. 2011), right when $<\! R_{\rm e}\!>$ is most
rapidly increasing (!), and then starts to drop at lower
redshifts. Thus, between $z\sim 2.5$ and $\sim 1$ the distribution of
effective radii of newly formed PEGs is rapidly evolving, and does so
in such a way that the birth rate of {\it normal size} PEGs is higher
than a sill increasing birth rate of compact ones. It is only below
$z\sim 1$ that the number density of compact PEGs starts to drop, and
some growth of individual galaxies must take place. Thus,
understanding the redshift evolution of $<\! R_{\rm e}\!>$ requires to
follow {\it simultaneously} the complex interplay between the growth
of individual galaxies {\it and} the birth of new PEGs and their size
distribution.  A key issue that remains to be settled concerns the
number density of compact PEGs in the local Universe: how does it
compare with the number density at redshift 1 or 2?

\section{The Specific Star Formation Rate}

The specific star formation rate (sSFR) of {\it main sequence} galaxies as a
function stellar mass and cosmic time, conveniently parameterized as sSFR$(M_*,t)\propto
M_*^\beta t^{-\gamma}$, plays a pivotal role in galaxy evolution.  It controls
the growth rate of galaxies, the evolution of their mass function, and
can have a direct effect even on quenching of star formation
itself. Despite its importance, the precise value of $\beta$ is still
uncertain, with values in the literature spanning a very wide range
($-0.4\lsim\beta\lsim 0$), the result depending on how star
forming galaxies are selected, and how SFR and $M_*$ are measured.

There is clear evidence that the sSFR drops systematically with time
since $z\sim 2$, by at least a factor of $\sim 20$, hence  $\gamma\simeq 2.2-2.5$. This drop with time (increase
  with lookback time and redshift) runs almost parallel to the
  specific mass increase rate of merging dark matter halos,
  thus hinting for a direct link between star formation rate and
  accretion rate of the haloes. A  simple
  interpretation of such quasi-parallelism is that the supply of fresh
  baryons (gas) feeding star formation in galaxies follows the dark matter
  halo accretion rate, as if baryons and dark matter were fairly bound to
  each other. However, beyond redshift $\sim 2$ the
  sSFR appears to flatten, then remaining nearly constant (i.e., $\gamma\simeq 0$) all the way
  to very high redshifts (e.g., Gonzalez et al. 2010). On the contrary,  the
  specific increase rate of dark matter haloes keeps increasing.

  This divergence from parallelism beyond $z\sim 2$ has caused some
  concern on the theoretical side (e.g., Weinmann, Neinstein \& Dekel
  2011), as it would apparently demand drastic modifications to widely
  adopted assumptions, as if baryons and dark matter were
  substantially decoupled at early cosmic times.  Thus, measuring the
  actual value of $\gamma$ for redshifts beyond $\sim 2.5$ has
  potentially strong implications for the relative behavior of baryons
  and dark matter.  It may also be especially relevant for
  reionization, but may be relatively unimportant for the actual mass
  growth of galaxies, which most takes place at lower redshifts.  In
  phenomenological models of galaxy evolution (e.g., Peng et
  al. 2010), compared to a $\gamma=0$ case a $\gamma > 0$ beyond
  $z\sim 2.5$ would imply a different ratio of the final galaxy
  stellar mass to the mass of the initial seed, having otherwise rather
  irrelevant effect on the final outcome.  But a precise measurement
  of the run of the sSFR with mass and time (i.e., of $\beta$ and
  $\gamma$) remains a central issue for a proper understanding of
  galaxy formation and evolution.  Contrary to previous results, it
  has been reported at this meeting that the sSFR may still be
  increasing somewhat even well beyond $z\sim 2$ (e.g., Stark et
  al. 2012). Thus, the issue remains unsettled and worth attracting
  further studies.

\begin{figure}[b]
% \vspace*{-2.0 cm}
\begin{center}
 \includegraphics[width=4in]{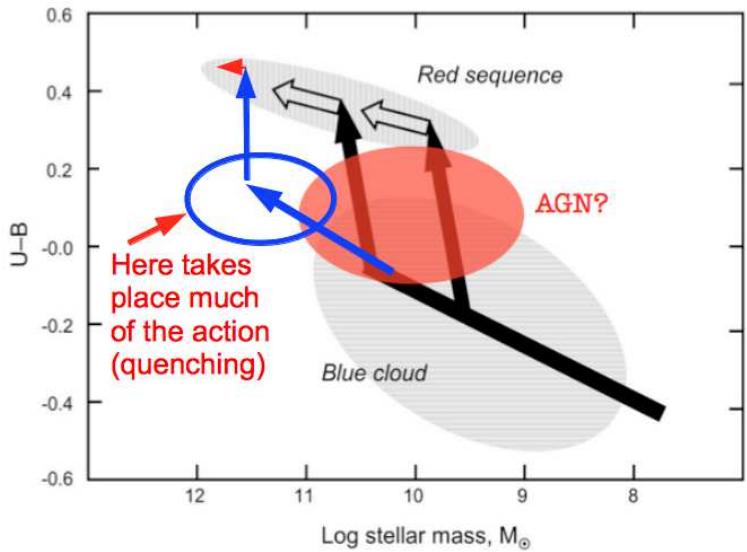} 
% \vspace*{-1.0 cm}
 \caption{A modification of the original cartoon  with an idealized sketch of the transition from star-forming ``blue cloud" galaxies to quenched ``red sequence" galaxies
 (black and grey scale, Faber et al. 2007). In color the actual path of galaxies being {\it mass quenched} is indicated (in blue), together with a modest further mass increase by dry merging (short red arrow).}
   \label{fig1}
\end{center}
\end{figure}

\section{Quenching and Dry Merging}

Quenching of star formation then turning to passive evolution is
perhaps the most salient event in the life of a galaxy.  Observations
tell us that the fraction of quenched galaxies is a strong function of
both stellar mass and local overdensity. Thus, two distinct physical
processes must exists, dubbed {\it mass quenching} and {\it
  environment quenching}, that act independent of each other (Peng et
al. 2010).  What remains to be established is the physical nature of
these two processes: what is environment quenching?  and, what is mass
quenching?

Is environment quenching just ram pressure stripping? or {\it
  strangulation}? or a combination thereof?  Is mass quenching a
process internal to galaxies themselves, such as e.g., AGN feedback?
or is it an external process, also related to the environment? (e.g.,
galaxies are quenched when the mass of the host dark matter halo
exceeds a threshold value).  Does mass quenching work by ejecting gas
{\it out of} galaxies or by preventing accretion of cold gas {\it
  into} galaxies?  In other words, what is {\it mass} in mass
quenching? Is it $M_*$ or $M_{\rm h}$?  Theoreticians appear to be
equally divided, and probably only observations can answer the
question.

\begin{figure}[b]
% \vspace*{-2.0 cm}
\begin{center}
 \includegraphics[clip,  scale=0.55]{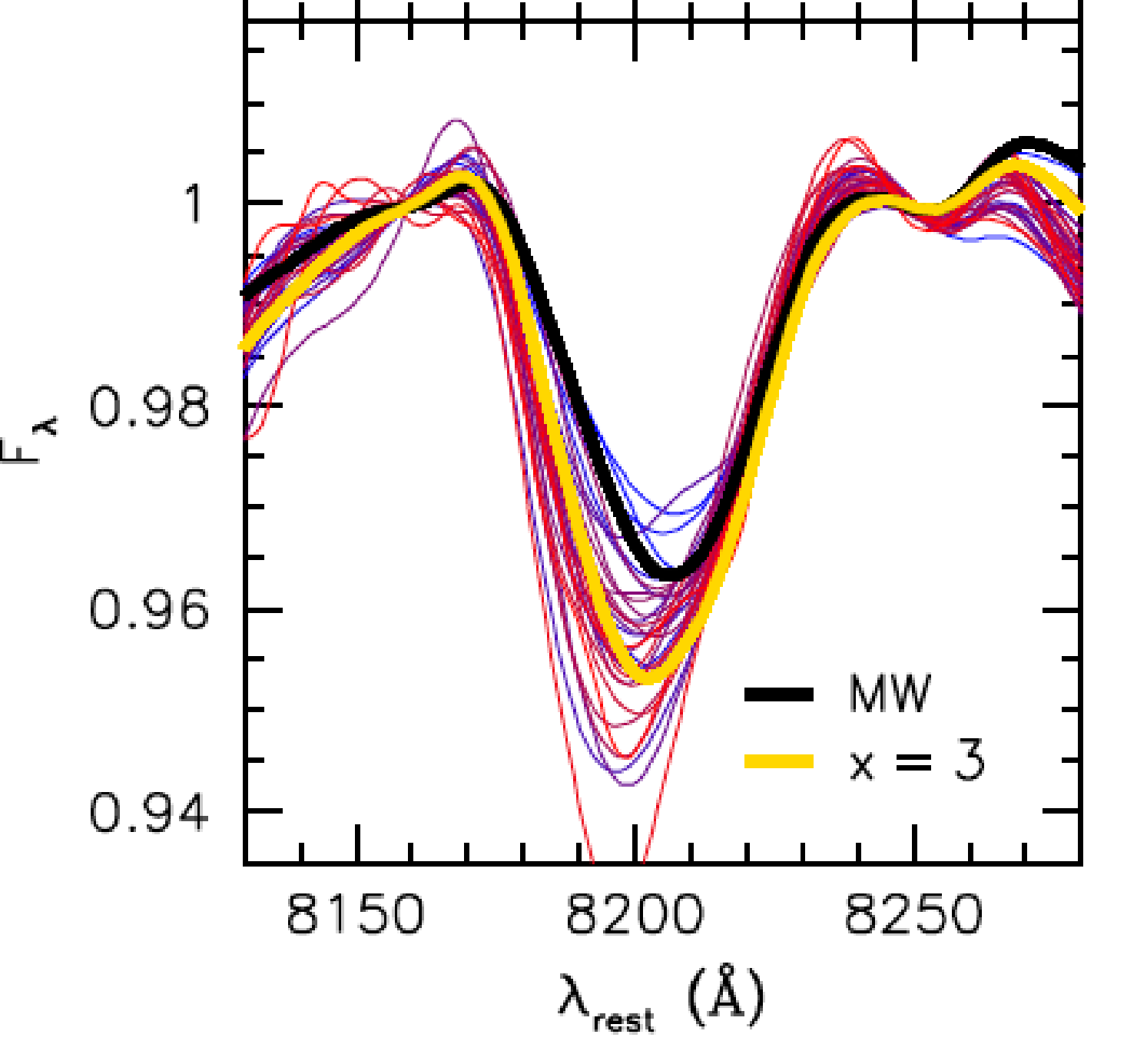} 
% \vspace*{-1.0 cm}
 \caption{The strength of the NaI $\lambda$8190 \AA\ doublet (here
   blended by velocity dispersion) in elliptical galaxies,
color
   coded blue to red with increasing velocity dispersion (from van
   Dokkum \& Conroy 2012). Two synthetic
stellar population spectra
   are also shown for an age of 13 Gyr and Milky Way bottom light IMF
   ($s = 1.35)$ and
a bottom heavy IMF ($s = 4$ equivalent to a
   logarithmic slope $x = 3$ as $s=x+1$).}
   \label{fig1}
\end{center}
\end{figure}

Yet, quenching may not be the last event in the evolution of massive
galaxies. Post-quenching (dry) merging can also take place, and in a
popular cartoon such further mass increase may be as large as a factor
$\sim 10$ (see Figure 1).  This scenario was motivated by the
perception that star-forming (SF) galaxies as massive as the most
massive quenched galaxies would not exist, hence massive PEGs
($M_*\gsim 10^{11}\,\msun$) could not form by just quenching star
formation in a {\it blue cloud} galaxy, then moved to the {\it red
  sequence}, i.e., there would be not enough massive blue galaxies in
the distant Universe with masses comparable to those of the brightest
red galaxies. Actually, massive enough SF galaxies do exist, but they
are very rare. The reason why they are rare is precisely because they
are growing in mass so rapidly that their life expectancy (as star
forming) is very short: they soon are going to be {\it mass quenched}!
Thus, the cartoon is missing the main point: most of the action,
i.e., most of the {\it mass quenching}, takes place near the top end
of the mass function of SF galaxies, for $M_*\gsim 10^{11}\,\msun$,
within the open ellipse in Figure 1, a region left empty in the
original plot.

Now, dry merging does certainly exist, but its role is not as prominent as sometimes envisaged.  
In high density regions  the Schechter $M_*^*$ of  quenched galaxies is just $\sim 0.1$ dex 
higher than in low density regions, where merging is almost absent (Peng et al. 2010). 
Thus, dry mergers make just an average $\sim 20\%$ mass increase, certainly not a factor $\sim 10$.

\section{The IMF}
Finally the IMF (usually described by a power law $dN\propto M^{-s}dM$).  It is currently common practice to
adopt a
universal IMF for a broad range of astrophysical applications, such as
e.g., to estimate stellar masses and
star formation rates of galaxies
from the local to the most distant Universe so far explored.  Yet,
it is perfectly legitimate to entertain the notion that the IMF may
not be universal. Indeed, from time to time one appeals to different IMFs to ease perceived
discrepancies between some theoretical
models and observations
(e.g., Dav\'e 2008), or between the dynamical and stelllar
population
mass to light ratios of galaxies (e.g., Cappellari et
al. 2012). Thus, sometimes one appeals to
a top heavy IMF with a lot
of massive stars boosting the luminosity and metal production rate,
sometime to a
bottom heavy IMF, with a lot of low mass stars, making
just mass but little light and no metals.
In particular, in a recent
surge of papers observational evidence has been presented that would
favor a very
bottom heavy IMF in massive elliptical galaxies. This
was based on the strength of the Na I $\lambda$8190 \AA\ doublet
and
the FeH Wing-Ford band at $\lambda$9900 \AA\ which both are strong in
dwarfs and weak in giants (van Dokkum \&
Conroy, 2012; Ferreras et
al. 2012; and references therein). These features appear to be
stronger
in such galaxies than in synthetic stellar population models which
adopt a Milky Way (bottom light) IMF, as
illustrated in Figure 2
(from van Dokkum \& Conroy 2012).

Notice that the depth of the NaI feature is just a few percent of the
(pseudo)continuum, and then a drastic variation of the IMF (with the
low-mass slope changing from s = 1.35 to 4) leads to a variation of the
central line depth from $\sim 96.2\%$ to just $\sim 95.2\%$. Formally,
as shown in Figure 1, even steeper IMFs would be required for the
ellipticals with the highest velocity dispersion. Clearly, the NaI
feature is very insensitive to the IMF slope and, moreover, deriving such
slope from the strength of this feature rests entirely on the
reliability of the stellar population models used to draw synthetic
spectra such as those shown in Figure 1. Thus, to trust the resulting
IMF one has to trust the ability of synthetic models to reproduce the
feature with exquisite accuracy. Therefore, a check of such models
should be mandatory before taking into serious consideration
systematic IMF variations as a function of the velocity dispersion of
elliptical galaxies. This is especially true given that models are
particularly uncertain (i.e., uncalibrated) for the super-solar
metallicities typical of the most massive elliptical galaxies.

Any inference on the IMF from integrated light ultimately rests on
synthetic stellar population models and therefore on their
reliability.  For example, Cappellari et al. (2012) have convincingly
demonstrated that mass closely follows light in the core of elliptical
galaxies, thus contributing to break the IMF-dark matter
degeneracy. Having constrained as marginal the dynamical effect of
dark matter, they find that in the core of elliptical galaxies the
$M/L$ ratio as inferred from dynamical modeling increases with $M_*$
and central velocity dispersion $\sigma$ more than current stellar
population models predict. They then interpret this departure in terms of a
systematic trend of the IMF with increasing galaxy mass and/or
$\sigma$. The $M/L$ ratio is certainly a more fundamental, bulk
property of galaxies compared to a weak spectral feature. Yet,
also in this case the inference on the IMF is correct only as long as  stellar
population models are correct. Worth noting is indeed that also
metallicity increases with mass and/or $\sigma$ and gets super solar
at high mass and $\sigma$: right where both the heavier IMF is
demanded, and where population models are less reliable. There is in
fact another degeneracy yet to be broken, one between the IMF and
stellar population models, whose use is unavoidable to infer anything
about the IMF.

\end{document}